\documentclass[]{spie}  

\usepackage{amsmath,amsfonts,amssymb}
\usepackage{graphicx}
\usepackage[colorlinks=true, allcolors=blue]{hyperref}
\usepackage{booktabs}
\usepackage{multicol}
\usepackage{multirow}
\usepackage{subcaption}

\newcommand\blfootnote[1]{%
  \begingroup
  \renewcommand\thefootnote{}\footnote{#1}%
  \addtocounter{footnote}{-1}%
  \endgroup
}

\title{A new dataset and comparison for multi-camera frame synthesis}

\author{Conall Daly}
\author{Anil Kokaram}
\affil{Sigmedia Group,  Electronic and Electrical Engineering Dept., Trinity College Dublin, \\ Dublin, Ireland}

\begin{document} 
\maketitle
\blfootnote{For further information contact Conall Daly: \href{mailto:dalyc21@tcd.ie}{dalyc21@tcd.ie}}
\begin{abstract}
Many methods exist for frame synthesis in image sequences but can be broadly categorised into frame interpolation and view synthesis techniques. Fundamentally, both frame interpolation and view synthesis tackle the same task, interpolating a frame given surrounding frames in time or space. However, most frame interpolation datasets focus on temporal aspects with single cameras moving through time and space, while view synthesis datasets are typically biased toward stereoscopic depth estimation use cases. This makes direct comparison between view synthesis and frame interpolation methods challenging. In this paper, we develop a novel multi-camera dataset using a custom-built dense linear camera array to enable fair comparison between these approaches. We evaluate classical and deep learning frame interpolators against a view synthesis method (3D Gaussian Splatting) for the task of view in-betweening. Our results reveal that deep learning methods do not significantly outperform classical methods on real image data, with 3D Gaussian Splatting actually underperforming frame interpolators by as much as 3.5 dB PSNR. However, in synthetic scenes, the situation reverses—3D Gaussian Splatting outperforms frame interpolation algorithms by almost +5 dB PSNR at a 95\% confidence level.
\end{abstract}


\section{Introduction}
\label{sec:intro}
The term {\em frame interpolation} typically refers to the process of synthesising new frames between existing frames in an image sequence, while {\em view synthesis} refers to the related task of generating images at new viewpoints given images from cameras positioned around a scene. Generally, frame interpolation involves the temporal interpolation between frames recorded from a single camera, whereas view synthesis involves the spatial interpolation between multiple camera viewpoints. These techniques have found diverse applications across multiple domains. Frame interpolation finds extensive application in video coding \cite{han_1997}, frame rate up-conversion \cite{castagno_96}, and video restoration \cite{werlberger2011optical,hilman_2000}. View synthesis has been used extensively as the cornerstone of image-based rendering in motion picture VFX \cite{debevec_98}. In practice, these two classes of techniques are often employed together in post-production, particularly in specialised effects such as TimeSlice and BulletTime \cite{kokaram_motion-based_2020}, where synthesis of frames in both space and time is required. While frame interpolation and view synthesis are applicable to very different scenarios, in the case of narrow baseline densely sampled scenes these approaches become comparable. In this paper we explore this particular use case.

In recent years, algorithms in both domains have transitioned from classical approaches, which rely on explicit understanding of motion and scene geometry, to advanced deep learning methodologies. This transition is primarily driven by deep learning models' abilities to learn complex, non-linear patterns and temporal dynamics from data that cannot be explicitly modelled. Around 2020, results from these modern techniques began to surpass the performance of traditional methods in terms of both accuracy and visual quality \cite{kokaram_motion-based_2020, kokaram_smpte}.

View synthesis applies not only to real image sequences captured by cameras but also to synthetic content such as animated films and games. In this context, view synthesis allows rendering engines to generate fewer shots for a scene and temporally upsample this to a desired frame rate. This makes synthetic content generation a prime research direction for these algorithms.

A notable challenge in advancing frame interpolation and view synthesis research is the scarcity of specialised datasets for multi-camera setups. Unlike conventional video datasets that focus on temporal interpolation from a single perspective \cite{su_2017}, multi-camera interpolation requires data that captures the same instant from multiple viewpoints. Creating such datasets requires ground truth at many possible in-between locations, similar to TimeSlice/BulletTime effects, requiring a dense array of cameras recording a scene simultaneously \cite{kettern_2010}.

In this paper, we address this issue by presenting a new densely sampled multi-view dataset created using our custom-built camera rig constructed with relatively low-cost components. We provide spatially and temporally smooth ground truth that can be used for both evaluation and training of algorithms for view in-betweening tasks. We then use this dataset to analyse a selection of frame interpolation and view synthesis algorithms chosen to represent broad classes of approaches, including Bayesian methods \cite{kokaram_bayesian_2020}, deep neural network (DNN) filter estimation techniques \cite{niklaus_video_2017-1, cheng_2022}, and radiance field scene rendering methods \cite{mildenhall_2020,kerbl_2023}.

\subsection{Contributions of this Paper}
Our main contributions are:
\begin{itemize}
    \item A new publicly available dataset for evaluating and training algorithms for view in-betweening
    \item A detailed analysis of contemporary video frame interpolation (VFI) and view synthesis algorithms using both real and synthetic datasets
\end{itemize}

\section{Background}

In both frame interpolation and view synthesis we are interested in generating an estimate ${\hat{I}}_{t+\Delta t}$ of a ground truth image ${{I}}_{t+\Delta t}$ at some time instant or location ${t+\Delta t}$. That location is in-between two pictures at $t$ and $t+1$ for example. All techniques, except 3D Gaussian Splatting \cite{kerbl_2023}, generate ${\hat{I}}_{t+\Delta t}$ given existing frames from ${{I}}_{t-n}$) to ${{I}}_{t+n}$ where $n$ is typically 1 but can be as much as 2 \cite{yu_2022, danier_2022}.

In this section we examine the three broad categories of frame interpolation and view synthesis algorithms that we evaluate using our novel dataset.

\subsection{Classical Methods}

By far the most popular and foundational framework for VFI is motion-compensated VFI which has been around since the late 1980’s. Thoma et al. \cite{thoma_motion_1989} developed the first motion-compensated de-interlacing VFI to take uncovering and occlusion of objects in the background of an image into account. This method outperformed more naive approaches of the time such as frame repetition and has become the dominant VFI framework. 

Some methods have extended motion compensated VFI by using Bayesian signal processing techniques. This approach allows for an algorithm designer to inject information about picture and motion smoothness through the use of prior probability distributions. These distributions typically take the form of  Markov random fields (MRFs).

The Bayesian approach to frame interpolation generalises many classical approaches. We  jointly estimate the frame we wish to interpolate $I_{t+\Delta t}$ and the interpolated motion field $F_{t+\Delta t}$. This is conditioned on the existing previous $I_t$ and $I_{t+1}$ future frames as well as forward $F_t$ and backward $F_{t+1}$ motion fields at those frames. The posterior distribution to be maximised can then be specified as follows.
\begin{equation}
    P(\hat{I}_{t+\Delta t}, \hat{F}_{t+\Delta t} | I_t, F_t, I_{t+1}, F_{t+1}) \propto P_{Image} \cdot P_{Motion}
\end{equation}
where the posterior is factorised into a likelihood $P_{Image}$ and prior $P_{Motion}$. The likelihood is usually a Gaussian distribution associated with the motion compensated error between the interpolated frame and existing frames at $t$ and $t+1$ given the motion information to be interpolated. The prior $P_{Motion}$ imposes smoothness on the motion fields.
The key to this approach is the exact specification of the priors followed by the conversion of the problem from probability maximisation to energy minimisation by a simple logarithmic transformation.

In this work we follow the the Bayesian frame interpolation technique proposed by Kokaram et al. (ACKMRF) \cite{kokaram_bayesian_2020, kokaram_motion-based_2020}. See that paper for details of the distributions and final algorithm. A version of this technique was implemented as the Kronos re-timer\footnote{\url{https://learn.foundry.com/nuke/content/reference_guide/time_nodes/kronos.html}} provided in Nuke. The only difference is that Kronos uses a robust initial estimate for the interpolated motion field provided by the OFlow motion estimator.

\subsection{Deep Learning Methods}
One of the earliest and most successful deep learning based frame interpolation methods is based on the idea of using a DNN to estimate filter kernels for image interpolation. These are known as dynamic filter networks \cite{vangool_2016}. Niklaus et al. \cite{niklaus_video_2017-1} used this approach for estimating the interpolated pixel value at a site $\mathbf{x}$ as follows.

\begin{equation}
\label{eq:niklaus-filters}
	\hat{I}_{t+\Delta t}(\mathbf{x})
	= 
	k_t(\mathbf{x}) * I_t(\mathbf{x})
	+
	k_{t+1}(\mathbf{x}) * I_{t+1}(\mathbf{x})
\end{equation}

Here two spatially varying filter kernels ($k_t$, $k_{t+1}$) are used to generate $\hat{I}_{t+\Delta t}(\mathbf{x})$ by convolution with the appropriate existing frame. In the original work, motion between frames was implicitly accounted for by the spatial extent of the kernel. Later on subsequent work recognised that if the future and previous frames were motion compensated, the kernel sizes could be reduced significantly with a positive impact on computational load.

This neural network is trained using a perceptual quality metric. It is based on a weighted combination of the $\mathcal{L}_1$ distance between a ground truth and interpolated frame along with the $\mathcal{L}_1$ distance between perceptual features from the outputs of VGG16 layers for the same frames.

Niklaus' approach motion estimation and occlusion handling are implicit within the filter estimation network. However, many other neural network based frame interpolation methods define separate networks for handling motion estimation, feature extraction and tracking etc. For example, Bao et al. \cite{bao_2019} use the pre-trained PWC-Net \cite{sun_2018} as a component in their frame interpolation network. They also train a sub-network to estimate depth from the input frames which is used to detect occlusion and uncovering.

In this work we use a selection of these tools for our comparative analysis. Every selected algorithm, except those of Niklaus et al., use the idea of motion compensated interpolation in some way, with motion being generated by a separate DNN sub-network.

We use the previously mentioned Revisiting Adaptive Convolutions for Video Frame Interpolation (revisiting-sepconv) \cite{niklaus_2021}, along with Softmax Splatting for Video Frame Interpolation (softmax-splatting) \cite{niklaus_2020}. We also choose a set of algorithms which do not rely on filter kernel estimation, but instead use a U-Net with motion and image inputs, Asymmetric Bilateral Motion Estimation for Video Frame Interpolation (ABME) \cite{park_2021}, All-Pairs Multi-Field Transforms for Efficient Frame Interpolation (AMT) \cite{li_2023}.  In addition, we compare with  Deep Bayesian Video Frame Interpolation (DBVI) \cite{yu_2022} which combines Bayesian approaches with DNNs for estimating frame and motion likelihoods.

We choose two other techniques which use simple motion compensated frame interpolation followed by a DNN for recovering the lost picture details using post-processing modules, A Spatio-Temporal Multi-Flow Network for Frame Interpolation (ST-MFNet) \cite{danier_2022}, A Unified Pyramid Recurrent Network for Video Frame Interpolation (UPR-Net) \cite{jin_2023}.

We also compare with a transformer based technique which also uses explicit motion information, Video Frame Interpolation with Transformer (VFIformer) \cite{lu_2022}.

\subsection{3D Gaussian Splatting}

Novel view synthesis by learning a 3D representation of a scene has become a very active research area in recent years with the advent of new radiance field rendering techniques. The most well known example is neural radiance fields (NeRFs) \cite{mildenhall_2020} which learn a neural network representation of the shape and view-dependent appearance of a scene. 

Recently the field has moved away from neural methods as neural network scene representation were found to be inefficient in training compared to conventional computer graphics data structures. These data structures can provide embedded priors that reduce time spent on learning empty space and low detail areas of the scene \cite{yu2021plenoctrees,yu_2021}.

3D Gaussian Splatting \cite{kerbl_2023} is one of the most recent and successful of these techniques. However, it is worth noting that splatting as a volume rendering technique has been around since the 1990's \cite{westover_1991}. Starting from a point cloud generated using a structure-from-motion (SfM) technique we interpolate the appearance between each point in the cloud using 3D Gaussians $\mathcal{N}(x)$. Learning the appearance of the scene then consists of optimising the covariance matrix ($\Sigma$) and opacity ($\alpha$) parameters for each Gaussian, which is defined by Equation \ref{eq:gaussian}.

\begin{equation}
\label{eq:gaussian}
    \alpha\cdot\mathcal{N}(x) = \alpha\cdot\left[e^{-\frac{1}{2}x^{\intercal}\Sigma^{-1}x}\right]
\end{equation}

Kerbl et al. \cite{kerbl_2023} introduce a method of growing, shrinking and pruning Gaussians to properly construct high detail and low detail features in the scene. A key innovation that makes it possible for 3D Gaussian Splatting to render in real-time and train quickly is the introduction of a a fast differentiable rasteriser. The rasteriser allows for back-propagation of the loss measured in image space into the scene space where we optimise our 3D representation. This loss used for learning a scene is a weighted combination of $\mathcal{L}_1$ and structural dissimilarity (D-SSIM). We employ 3D Gaussian Splatting (gaussian-splatting) \cite{kerbl_2023} as a 3D scene reconstruction method to compare against our selection of frame interpolators.

\section{Proposed Dataset}

\begin{figure}
    \centering
    \begin{tabular}{cc}
    $\vcenter{\hbox{\includegraphics[width=0.4\linewidth]{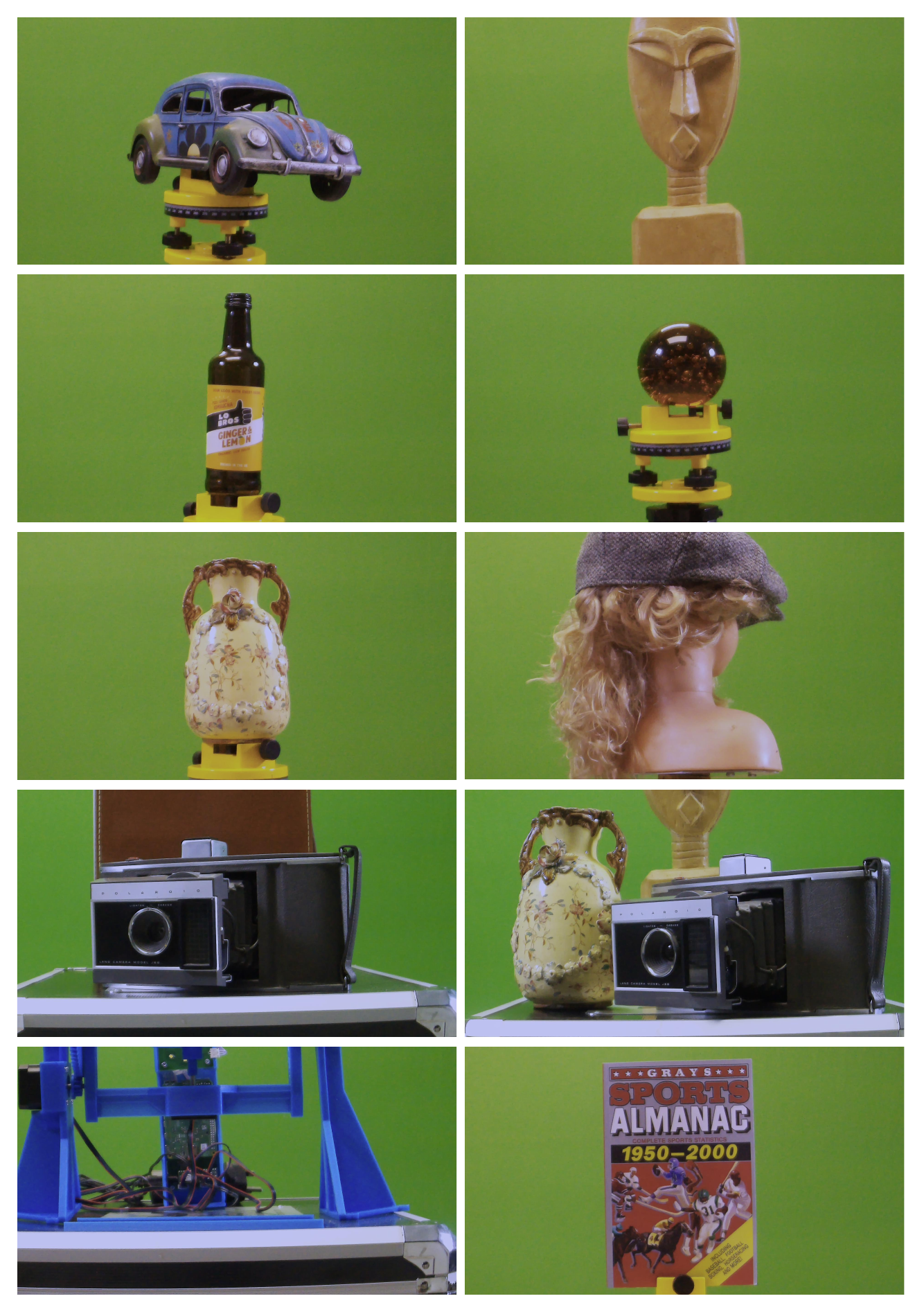}}}$  & $\vcenter{\hbox{\includegraphics[width=0.45\linewidth]{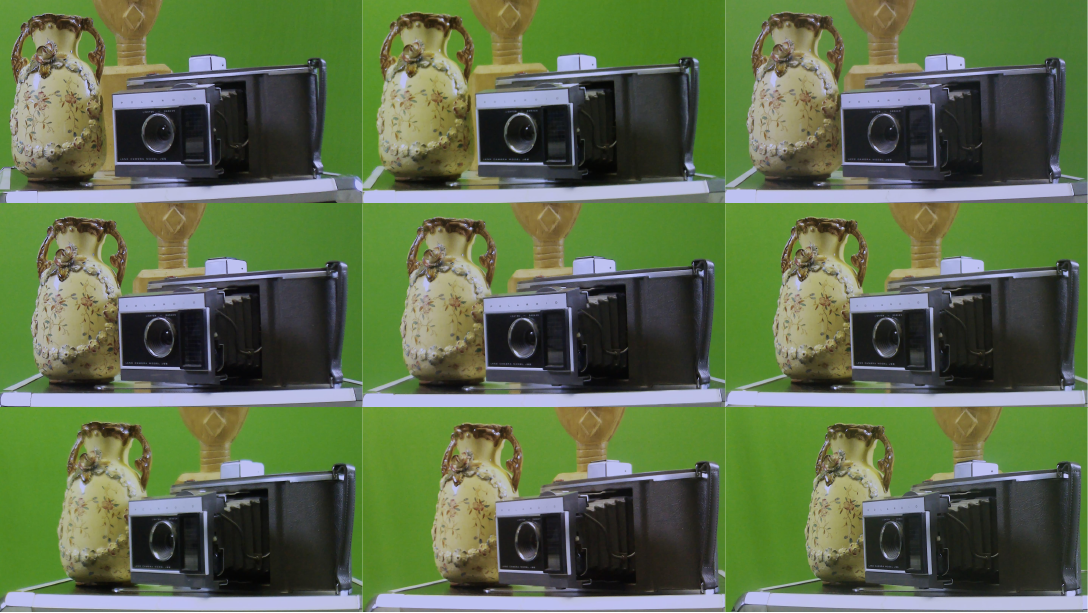}}}$ \\
     a)    &  b)
    \end{tabular}
    \caption{a) \textit{1080p images of real dataset objects at a medium
distance, captured using our multi-view rig}. b) \textit{Sweep through of a selected sequence in 1080p dataset showing parallax effect and magnitude of motion between frames, frames are ordered in increasing order left to right and then top to bottom (i.e. furthest left camera on rig is shown in top-left and furthest right camera on rig is shown in bottom-right)}.}
    \label{fig:dataset}
\end{figure}

Popular view interpolation datasets focus on temporal interpolation with single handheld cameras, ranging from consumer quality (Adobe240fps \cite{su_2017}) to professional (Netflix\footnote{\url{opencontent.netflix.com}}, Xiph4K\footnote{\url{media.xiph.org/video/derf/}}). Multi-view datasets are typically limited to dual-camera setups for depth estimation or scene reconstruction with insufficient spatial sampling and no frame ordering \cite{yu2023mvimgnet, geusebroek_2005}.

Our multi-view dataset, shown in Figure \ref{fig:dataset} addresses these limitations, providing evaluation-ready data for both frame interpolation and view synthesis algorithms. Our new dataset can be downloaded at the following link\footnote{Dataset download link: \url{https://drive.google.com/drive/folders/1J7QdGFcYw_AAAO6U9TNgBvTJwNCTYUo-?usp=sharing}}.

\subsection{Multi-view Capture Rig}

Our custom capture rig comprises a linear array of 9 Raspberry Pi units, each with a High Quality Camera\footnote{\url{www.raspberrypi.com/products/raspberry-pi-high-quality-camera/}} featuring a 12.3 Megapixel Sony IMX477 sensor. A controller PC interfaces through a network switch to provide synchronized capture and centralized storage. Camera spacing is constrained by the 38mm sensor board width\footnote{\url{datasheets.raspberrypi.com/hq-camera/hq-camera-cs-mechanical-drawing.pdf}}, resulting in approximately 38mm center-to-center distances across a 34.2cm total span. Custom 3D-printed mounts accommodate the close spacing requirements.

We capture 10 objects (Figure \ref{fig:dataset}a) using two-point LED lighting against a green screen backdrop. Objects are recorded at three distances---close (0.75m), medium (1m), and far (1.25m)---to create varying parallax effects. Figure \ref{fig:dataset}b shows an example sequence assembled from consecutive left-to-right camera views.


\subsection{Post-Processing Pipeline}

Each camera records an image with a slightly different colour space. Furthermore the orientation of the camera plane is often not what was intended. When played back as a sequence, these views show very poor colour smoothness and heavy judder.  We therefore implement a number of post-processing steps to correct these distortions using a pipeline in Nuke\footnote{\url{www.foundry.com/products/nuke-family/nuke}}. First, we remove geometric distortion from each frame. The distortion is estimated by capturing a chequerboard grid with each camera and using grid detection and Nuke's LensDistortion\footnote{\url{learn.foundry.com/nuke/content/reference_guide/transform_nodes/lensdistortion.html}} node to estimate parameters for the distortion model.

Next, we perform a colour correction step. It is not necessary to estimate some "exact" colour reproduction of the scene, only to smooth out the differences between cameras as much as possible. Hence we then balance the colour across all cameras using a Calibrite ColorChecker Classic chart\footnote{\url{calibrite.com/de/product/colorchecker-classic}} paired with the CalibrateMacbeth Blinkscript node developed by Jedediah Smith\footnote{\url{gist.github.com/jedypod/798b365ea64e8121999e7036ae7e0217}}.

We then stabilise the frames using the Tracker node in Nuke\footnote{\url{learn.foundry.com/nuke/content/tutorials/written_tutorials/tutorial2/stabilizing_elements.html}}. Two points are selected in each scene and an affine warp is applied to account for movement of the image plane of the cameras. While general 3D scenes should be aligned using a projective transformation, an affine warp can be used in the case of linear camera arrays with close camera spacing.

Final processing crops scenes to 1080p resolution with 720p downsampled versions. We refer to these in the text as \emph{Real1080p} and \emph{Real720p} respectively.

\begin{figure*}
    \centering
    \includegraphics[width=0.8\linewidth]{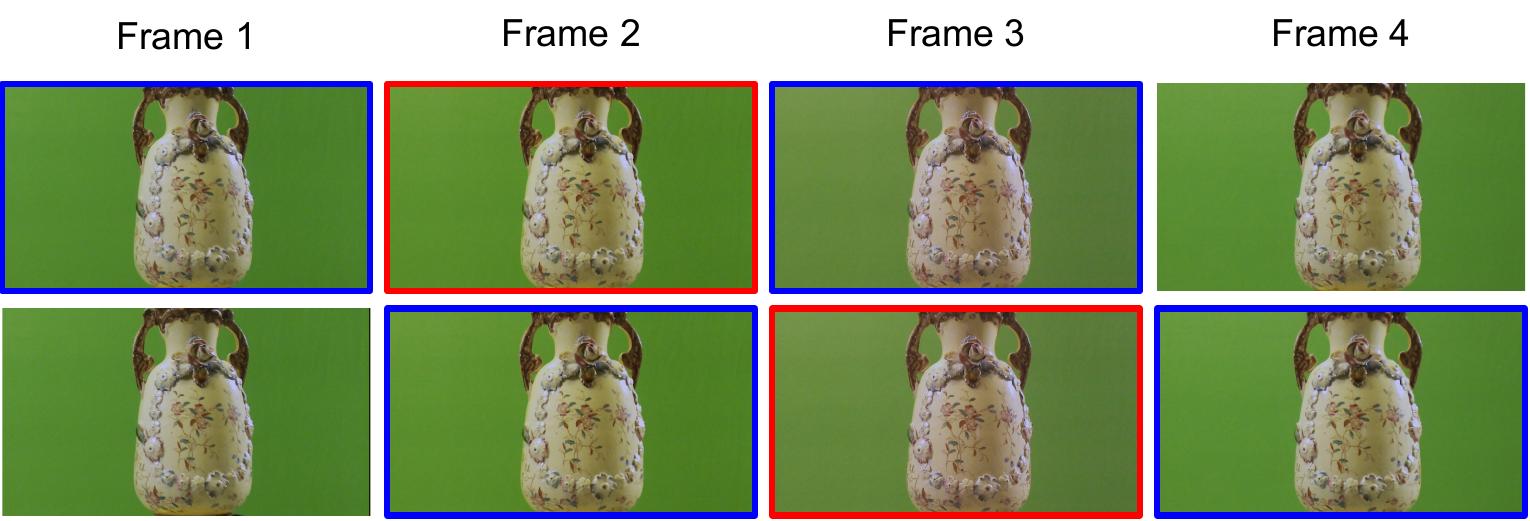}
    \caption{\textit{Input frames shown in {\color{blue} blue} and used to interpolate an estimate for the ground truth frame shown in {\color{red} red}. We slide this window from left to right across our sequence so $7$ interpolated frames are generated. We show here the leftmost $4$ frames in our $9$ camera array for one object. The top row shows the interpolation window to generate frame $2$, and the bottom row shows the next window in the sequence to generate frame $3$.}}
    \label{fig:vase}
\end{figure*}

\subsection{Synthetically Generated Views}

For synthetic data, we use training frames from Mildenhall et al. \cite{mildenhall_2020}, which provide ordered smooth camera paths suitable for both frame interpolators and view synthesis algorithms. This \emph{Synthetic} subset features 800$\times$800 resolution and contains 1,600 frames, while Real720p and Real1080p each contain 270 frames, totalling 2,140 frames across the complete dataset.

In essence the synthetic data presents a best case scenario for view synthesis, being noise free and free from blurring. This allows for easier feature extraction with COLMAP and less of challenge for frame interpolators.

\section{Evaluation Methodology}

Our evaluation methodology assesses each algorithm's ability to interpolate the middle frame from every triplet of consecutive views. Given cameras 1, 2, and 3, we interpolate view 2 using views 1 and 3, then slide this interpolation window across the sequence as illustrated in Figure \ref{fig:vase}. Since start and end frames cannot serve as ground truth, the maximum evaluable frames is 2,004.

Some algorithms (DBVI, ST-MFNet) require two frames before and after the interpolated frame. We ensure fair comparison by evaluating only where sufficient frames exist, algorithms are not penalized for these requirements when calculating metrics. Gaussian-splatting requires SfM preprocessing via COLMAP \cite{schoenberger_2016}. When this error-prone preprocessing fails to generate camera parameters, those cases are excluded from gaussian-splatting's averages for PSNR and SSIM.

Statistical significance testing evaluates whether observed differences are meaningful at the 95\% confidence level using bootstrap confidence intervals and significance tests\footnote{\url{https://sjeng.org/bootstrap.html}}. All algorithms run on an Nvidia A4000 workstation with 16GB VRAM. Out-of-memory (OoM) indicators mark algorithms that exceeded this memory limit during execution.

\section{Results and Discussion}

Table \ref{tab:res-tab} presents PSNR and SSIM statistics for each algorithm across dataset subsets. ST-MFNet and UPR-Net achieve the highest PSNR on Real720p and Real1080p respectively, though narrow mean spreads and large standard deviations suggest limited significance for Real720p data. Statistical testing confirms no significant differences between algorithms for Real720p ($p>0.05$). Complete p-values are provided in Appendix \ref{sec:p-val}.

\begin{table}[]
\centering

\caption{\textit{Mean ($\mu$) and standard deviation ($\sigma$) of PSNR (dB) and SSIM across all sequences for each subset of the dataset. The best and second best algorithm for each subset are \textbf{bolded} and \underline{underlined} respectively.}} 
\label{tab:res-tab}

\resizebox{\columnwidth}{!}{
\begin{tabular}{@{}c||cccc||cccc||cccc@{}}
\toprule
                   & \multicolumn{4}{c||}{Real720p}                                                                                                     & \multicolumn{4}{c||}{Real1080p}                                                                                                    & \multicolumn{4}{c}{Synthetic}                                                                                                    \\ \midrule
\multirow{2}{*}{}  & \multicolumn{2}{c|}{PSNR (dB)}                                  & \multicolumn{2}{c||}{SSIM}                                       & \multicolumn{2}{c|}{PSNR (dB)}                                  & \multicolumn{2}{c||}{SSIM}                                       & \multicolumn{2}{c|}{PSNR (dB)}                                  & \multicolumn{2}{c}{SSIM}                                       \\ \cmidrule(l){2-13} 
                   & \multicolumn{1}{c}{$\mu\uparrow$} & \multicolumn{1}{c|}{$\sigma$} & \multicolumn{1}{c}{$\mu\uparrow$} & \multicolumn{1}{c||}{$\sigma$} & \multicolumn{1}{c}{$\mu\uparrow$} & \multicolumn{1}{c|}{$\sigma$} & \multicolumn{1}{c}{$\mu\uparrow$} & \multicolumn{1}{c||}{$\sigma$} & \multicolumn{1}{c}{$\mu\uparrow$} & \multicolumn{1}{c|}{$\sigma$} & \multicolumn{1}{c}{$\mu\uparrow$} & \multicolumn{1}{c}{$\sigma$} \\ \midrule
ABME               & \underline{29.08}                             & \multicolumn{1}{l|}{3.78}   & \underline{0.897}                             & 0.055                       & \underline{28.51}                             & \multicolumn{1}{l|}{3.78}   & \textbf{0.892}                             & 0.05                        & 30.27                             & \multicolumn{1}{l|}{3.50}    & 0.960                              & 0.024                      \\
AMT                & 28.98                             & \multicolumn{1}{l|}{3.92}   & 0.896                             & 0.058                       & OoM                               & \multicolumn{1}{l|}{OoM}    & OoM                               & OoM                         & 29.68                             & \multicolumn{1}{l|}{3.55}   & 0.953                             & 0.031                      \\
DBVI               & 28.45                             & \multicolumn{1}{l|}{2.93}   & 0.894                             & 0.047                       & 22.68                             & \multicolumn{1}{l|}{1.88}   & 0.833                             & 0.041                       & 29.66                             & \multicolumn{1}{l|}{3.36}   & 0.934                             & 0.022                      \\
revisiting-sepconv & 28.61                             & \multicolumn{1}{l|}{4.30}    & 0.891                             & 0.065                       & 28.04                             & \multicolumn{1}{l|}{4.36}   & 0.888                             & 0.058                       & 28.55                             & \multicolumn{1}{l|}{3.53}   & 0.937                             & 0.038                      \\
softmax-splatting  & 28.70                              & \multicolumn{1}{l|}{3.87}   & 0.887                             & 0.061                       & 28.05                             & \multicolumn{1}{l|}{4.00}      & 0.879                             & 0.057                       & 29.89                             & \multicolumn{1}{l|}{3.46}   & 0.959                             & 0.027                      \\
ST-MFNet           & \textbf{29.96}                             & \multicolumn{1}{l|}{3.64}   & \textbf{0.908}                             & 0.050                        & OoM                               & \multicolumn{1}{l|}{OoM}    & OoM                               & OoM                         & 28.83                             & \multicolumn{1}{l|}{3.76}   & 0.924                             & 0.051                      \\
UPR-Net            & 29.02                             & \multicolumn{1}{l|}{3.77}   & 0.896                             & 0.056                       & \textbf{28.57}                             & \multicolumn{1}{l|}{3.70}    & \underline{0.890}                              & 0.052                       & \underline{30.51}                            & \multicolumn{1}{l|}{3.50}    & \underline{0.962}                             & 0.023                      \\
VFIformer          & 28.88                             & \multicolumn{1}{l|}{3.98}   & 0.896                             & 0.058                       & OoM                               & \multicolumn{1}{l|}{OoM}    & OoM                               & OoM                         & 30.28                           & \multicolumn{1}{l|}{3.66}   & 0.961                             & 0.025                      \\
ACKMRF             & 28.85                             & \multicolumn{1}{l|}{4.03}   & 0.883                             & 0.066                       & 28.35                             & \multicolumn{1}{l|}{4.07}   & 0.879                             & 0.06                        & 28.47                             & \multicolumn{1}{l|}{3.43}   & 0.935                             & 0.044                      \\
gaussian-splatting & 29.03                             & \multicolumn{1}{l|}{3.46}   & \textbf{0.908}                             & 0.038                       & 24.91                             & \multicolumn{1}{l|}{6.44}   & 0.847                             & 0.21                        & \textbf{34.82}                             & \multicolumn{1}{l|}{4.32}   & \textbf{0.978}                             & 0.017                      \\ \bottomrule
\end{tabular}}
\end{table}

Gaussian-splatting dominates synthetic data, outperforming UPR-Net by over 3.5 dB ($p=0.0$). However, on Real1080p data we notice a dramatic reduction by 10 dB. Recall that our PSNR measurements excludes cases where COLMAP fails completely.

\begin{figure}
\centering
\begin{tabular}{cc}
    \includegraphics[width=0.475\linewidth]{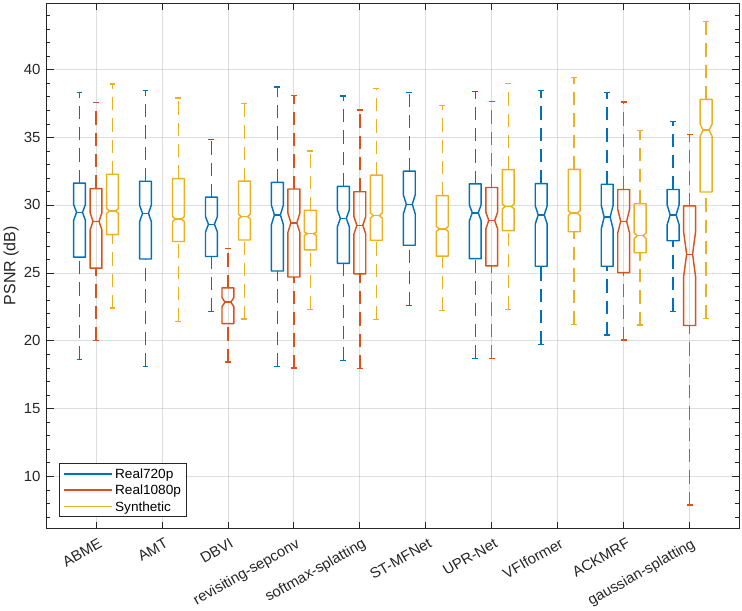} & \includegraphics[width=0.475\linewidth]{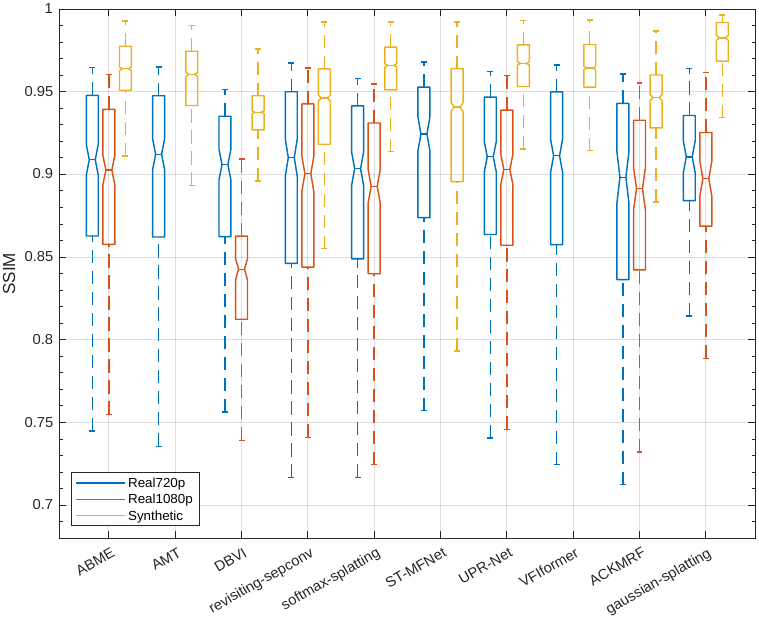} \\
         & 
    \end{tabular}
    \centering
    \caption{\textit{Box and whisker plots of PSNR (dB) (left) and SSIM (right) across each subset of the datasets.}}
    \label{fig:psnr_box_whisker}
\end{figure}

Figure \ref{fig:psnr_box_whisker} visualizes these performance differences through box-and-whisker plots. The notches provide visual hypothesis testing: non-overlapping notches indicate significantly different medians at the 5\% level. Gaussian-splatting exhibits extreme variability with PSNR ranging from 8 dB to 43.5 dB. The asymmetric distribution (wider bottom 50\% for Real720p and Synthetic) contrasts with more balanced distributions from other algorithms, highlighting gaussian-splatting's fragility when SfM preprocessing fails.

\begin{figure}
    \centering
    \includegraphics[width=0.95\linewidth]{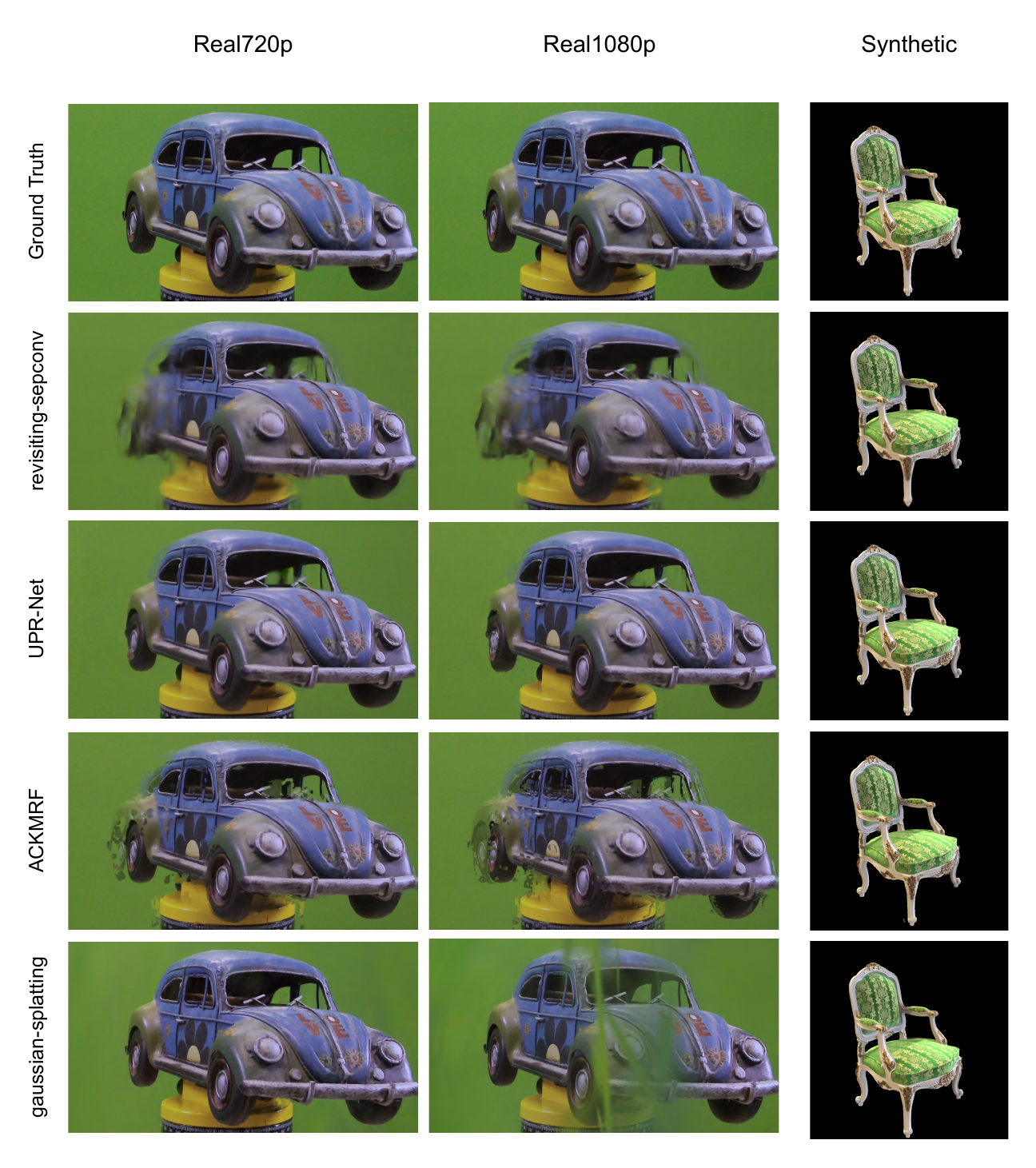}
    \caption{\textit{Comparison of synthesised frames for camera $4$ at a close distance ($0.75$m) across a selection of algorithms (right) and frame $136$ for synthetically rendered chair, the ground truth which we are comparing against is shown in the topmost row. In general all algorithms perform well on the synthetic dataset from a visual quality point of view. However,  the real dataset presents a challenge for all algorithms, with UPR-Net appearing to give the most consistent visually realistic result across all $3$ tests.}}
    \label{fig:01_01_04-chair_r_136}
\end{figure}

Figure \ref{fig:01_01_04-chair_r_136} shows visual comparisons across subsets. All methods perform well on synthetic data, with minor errors only visible on chair extremities closest to the viewer. Real data reveals the greatest algorithmic disparities: frame interpolators introduce warping artifacts (revisiting-sepconv) or disjointed patches (ACKMRF), while gaussian-splatting creates ``floater'' artifacts (stray Gaussians) that occlude scenes due to incorrect depth ordering during rasterisation. These large, incorrectly positioned Gaussians are particularly problematic in Real1080p data, demonstrating the reduced robustness of gaussian-splatting compared to frame interpolators.

SSIM results (Table \ref{tab:res-tab}, Figure \ref{fig:psnr_box_whisker}) show gaussian-splatting achieving highest scores for Real720p ($p\leq0.027$) and Synthetic ($p=0.000$) data. Its superior SSIM performance relative to PSNR likely stems from preserving 3D scene structure better than 2D-focused frame interpolators, maintaining object shapes even when artifacts appear.

Notably, deep learning frame interpolators do not significantly outperform classical Bayesian methods (ACKMRF) on real data. DBVI, a hybrid approach, performs significantly worse than ACKMRF on Real1080p data ($\approx$6 dB difference, $p=0.000$), highlighting deep learning's sensitivity to resolutions outside training distributions.

\begin{figure}
    \centering
    \includegraphics[width=0.6\linewidth]{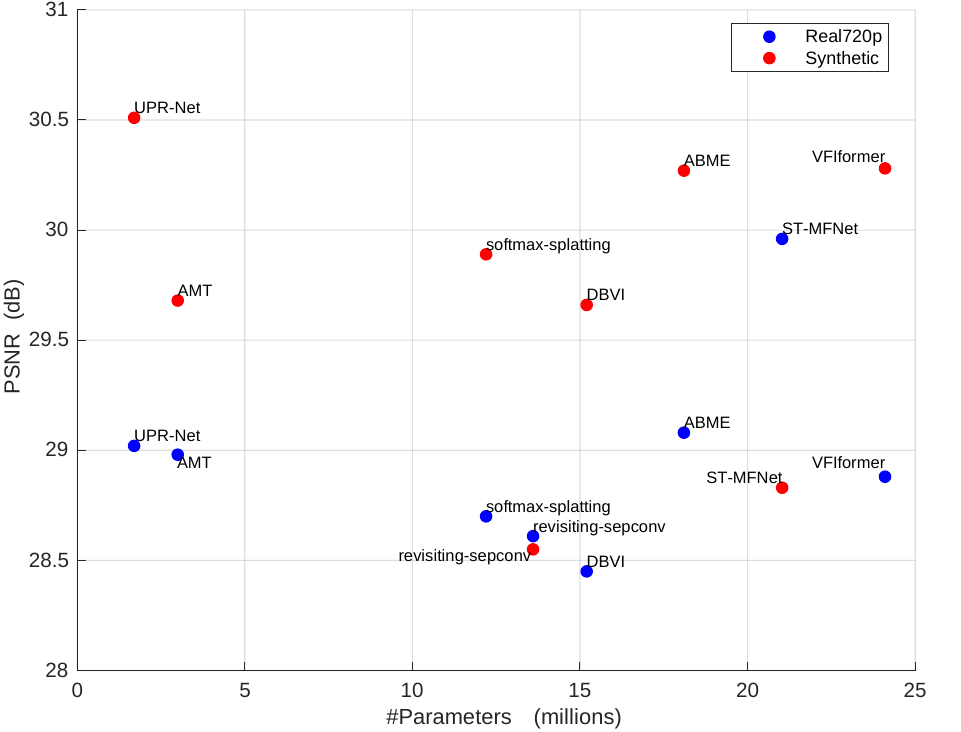}
    \caption{\textit{PSNR (dB) plotted against computational complexity of neural network measured in millions of parameters. Points which are in the top left of the graph are considered more parameter efficient, given the best performance of the smallest computational cost.}}
    \label{fig:psnr-comp-cmplx}
\end{figure}

Figure \ref{fig:psnr-comp-cmplx} examines computational efficiency by plotting PSNR versus neural network parameters. UPR-Net emerges as the most parameter-efficient for synthetic data, achieving top performance with $\approx$1 million parameters. However, UPR-Net lags behind ST-MFNet on Real720p data by $\approx$1 dB (possibly non-significant, $p=1$). ACKMRF requires 5 iterations per frame while gaussian-splatting needs $\approx$30,000 iterations for scene representation, placing them at opposite ends of the computational spectrum.

\section{Conclusions}

We present a unique comparison of classical and deep learning frame interpolators against state-of-the-art 3D scene view synthesis algorithms, enabled by a novel dataset containing 10 objects recorded from 3 depths across 9 cameras, plus synthetic data for fair comparison, yielding 2,140 frames for evaluating algorithms.

Our results reveal that 3D Gaussian Splatting excels on synthetic scenes, achieving nearly +5 dB PSNR improvement over frame interpolation algorithms with 95\% statistical confidence, but struggles with real-world image datasets where it performs up to 3.5 dB PSNR worse than frame interpolation methods. This performance disparity likely stems from preprocessing failures in the SfM pipeline, which works reliably only with synthetic data's sharp, well-behaved features but falters on real data's complexities.

When algorithms fail, they produce distinct artifacts that reveal their underlying mechanisms. 3D Gaussian Splatting preserves scene structure better than frame interpolators, which tend to introduce warping and patchy regions, but suffers from floating artifacts during rasterisation. Notably, classical Bayesian frame interpolation performs comparably to deep learning approaches, particularly on Real720p, offering similar results at lower computational cost, which suggests classical methods remain viable alternatives to more complex deep learning approaches.

Future work will expand to dynamic scenes with motion blur, requiring more extensive analysis of spatially and temporally moving footage across multiple viewpoints. We also plan to investigate whether alternative SfM algorithms could improve 3D Gaussian Splatting's robustness.

\acknowledgments 
 
This publication has emanated from research jointly funded by Taighde Éireann – Research Ireland, and Overcast HQ under Grant number EPSPG/2023/1515.

\bibliography{report}
\bibliographystyle{spiebib}

\newpage
\appendix

\section{P-values Obtained Using Bootstrapped Confidence Intervals}
\label{sec:p-val}

\begin{table}[h]
\centering
\caption{\textit{P-values for PSNR comparing the difference in means between an algorithm in the leftmost column with an algorithm in the topmost row. Statistically significant differences with 95\% confidence are marked with a *. The algorithms revisiting-sepconv (re-sepconv), softmax-splatting (soft-splat), and gaussian-splatting (g-splat) have been abbreviated for neater formatting.}}
\label{tab:p-val-psnr}
\scalebox{0.9}{
\begin{tabular}{@{}llllllllll@{}}
\toprule
\multicolumn{1}{c}{} & AMT    & DBVI   & re-sepconv & soft-splat & ST-MFNet & UPR-Net & VFIformer & ACKMRF & g-splat \\ \midrule
\multicolumn{10}{l}{\textbf{Real720p}}                                                                               \\ \midrule
ABME                 & 1      & 0.866  & 0.991      & 0.999      & 1        & 1       & 1         & 0.995  & 1       \\
AMT                  & -      & 0.978  & 0.999      & 1          & 1        & 1       & 1         & 1      & 0.999   \\
DBVI                 & -      & -      & 1          & 0.998      & 0.796    & 0.913   & 0.997     & 1      & 0.552   \\
re-sepconv           & -      & -      & -          & 1          & 0.982    & 0.995   & 1         & 1      & 0.933   \\
soft-splat           & -      & -      & -          & -          & 0.997    & 1       & 1         & 1      & 0.984   \\
ST-MFNet             & -      & -      & -          & -          & -        & 1       & 0.999     & 0.989  & 1       \\
UPR-Net              & -      & -      & -          & -          & -        & -       & 1         & 0.997  & 1       \\
VFIformer            & -      & -      & -          & -          & -        & -       & -         & 1      & 0.993   \\
ACKMRF               & -      & -      & -          & -          & -        & -       & -         & -      & 0.948   \\ \midrule
\multicolumn{10}{l}{\textbf{Real1080p}}                                                                              \\ \midrule
ABME                 & OoM    & 0.000* & 0.939      & 0.94       & OoM      & 0.999   & OoM       & 0.949  & 0.020*  \\
AMT                  & -      & OoM    & OoM        & OoM        & OoM      & OoM     & OoM       & OoM    & OoM     \\
DBVI                 & -      & -      & 0.000*     & 0.000*     & OoM      & 0.000*  & OoM       & 0.000* & 0.267   \\
re-sepconv           & -      & -      & -          & 0.999      & OoM      & 0.92    & OoM       & 0.998  & 0.139   \\
soft-splat           & -      & -      & -          & -          & OoM      & 0.923   & OoM       & 0.999  & 0.092   \\
ST-MFNet             & -      & -      & -          & -          & -        & OoM     & OoM       & OoM    & OoM     \\
UPR-Net              & -      & -      & -          & -          & -        & -       & OoM       & 0.94   & 0.015*  \\
VFIformer            & -      & -      & -          & -          & -        & -       & -         & OoM    & OoM     \\
ACKMRF               & -      & -      & -          & -          & -        & -       & -         & -      & 0.087   \\ \midrule
\multicolumn{10}{l}{\textbf{Synthetic}}                                                                              \\ \midrule
ABME                 & 0.021* & 0.014* & 0.000*     & 0.19       & 0.000*   & 0.737   & 0.995     & 0.000* & 0.000*  \\
AMT                  & -      & 0.995  & 0.017*     & 0.802      & 0.034*   & 0.000*  & 0.020*    & 0.000* & 0.000*  \\
DBVI                 & -      & -      & 0.017*     & 0.783      & 0.034*   & 0.000*  & 0.014*    & 0.000* & 0.000*  \\
re-sepconv           & -      & -      & -          & 0.001*     & 0.993    & 0.000*  & 0.000*    & 0.631  & 0.000*  \\
soft-splat           & -      & -      & -          & -          & 0.003*   & 0.008*  & 0.18      & 0.000* & 0.000*  \\
ST-MFNet             & -      & -      & -          & -          & -        & 0.000*  & 0.000*    & 0.52   & 0.000*  \\
UPR-Net              & -      & -      & -          & -          & -        & -       & 0.783     & 0.000* & 0.000*  \\
VFIformer            & -      & -      & -          & -          & -        & -       & -         & 0.000* & 0.000*  \\
ACKMRF               & -      & -      & -          & -          & -        & -       & -         & -      & 0.000*  \\ \bottomrule
\end{tabular}}
\end{table}

\begin{table}[h]
\centering
\caption{\textit{P-values for SSIM comparing the difference in means between an algorithm in the leftmost column with an algorithm in the topmost row. Statistically significant differences with 95\% confidence are marked with a *. The algorithms revisiting-sepconv (re-sepconv), softmax-splatting (soft-splat), and gaussian-splatting (g-splat) have been abbreviated for neater formatting.}}
\label{tab:p-val-ssim}
\scalebox{0.9}{
\begin{tabular}{@{}llllllllll@{}}
\toprule
           & AMT    & DBVI   & re-sepconv & soft-splat & ST-MFNet & UPR-Net & VFIformer & ACKMRF & g-splat \\ \midrule
\multicolumn{10}{l}{\textbf{Real720p}}                                                                     \\ \midrule
ABME       & 1      & 0.000* & 0.889      & 0.85       & 0.000*   & 1       & 1         & 0.293  & 0.027*  \\
AMT        & -      & 0.000* & 0.941      & 0.931      & 0.000*   & 1       & 1         & 0.419  & 0.020*  \\
DBVI       & -      & -      & 0.000*     & 0.000*     & 1        & 0.000*  & 0.000*    & 0.000* & 0.000*  \\
re-sepconv & -      & -      & -          & 1          & 0.000*   & 0.941   & 0.947     & 0.947  & 0.001*  \\
soft-splat & -      & -      & -          & -          & 0.000*   & 0.928   & 0.941     & 0.947  & 0.000*  \\
ST-MFNet   & -      & -      & -          & -          & -        & 0.000*  & 0.000*    & 0.000* & 0.000*  \\
UPR-Net    & -      & -      & -          & -          & -        & -       & 1         & 0.405  & 0.016*  \\
VFIformer  & -      & -      & -          & -          & -        & -       & -         & 0.451  & 0.018*  \\
ACKMRF     & -      & -      & -          & -          & -        & -       & -         & -      & 0.000*  \\ \midrule
\multicolumn{10}{l}{\textbf{Real1080p}}                                                                    \\ \midrule
ABME       & OoM    & 0.000* & 0.88       & 0.531      & OoM      & 0.896   & OoM       & 0.336  & 0.437   \\
AMT        & -      & OoM    & OoM        & OoM        & OoM      & OoM     & OoM       & OoM    & OoM     \\
DBVI       & -      & -      & 0.000*     & 0.000*     & OoM      & 0.000*  & OoM       & 0.000* & 0.000*  \\
re-sepconv & -      & -      & -          & 0.896      & OoM      & 0.896   & OoM       & 0.768  & 0.582   \\
soft-splat & -      & -      & -          & -          & OoM      & 0.618   & OoM       & 0.896  & 0.706   \\
ST-MFNet   & -      & -      & -          & -          & -        & OoM     & OoM       & OoM    & OoM     \\
UPR-Net    & -      & -      & -          & -          & -        & -       & OoM       & 0.43   & 0.46    \\
VFIformer  & -      & -      & -          & -          & -        & -       & -         & OoM    & OoM     \\
ACKMRF     & -      & -      & -          & -          & -        & -       & -         & -      & 0.774   \\ \midrule
\multicolumn{10}{l}{\textbf{Synthetic}}                                                                    \\ \midrule
ABME       & 0.008* & 0.000* & 0.000*     & 0.911      & 0.000*   & 0.776   & 0.916     & 0.000* & 0.000*  \\
AMT        & -      & 0.000* & 0.000*     & 0.119      & 0.000*   & 0.000*  & 0.001*    & 0.000* & 0.000*  \\
DBVI       & -      & -      & 0.916      & 0.000*     & 0.129    & 0.000*  & 0.000*    & 0.916  & 0.000*  \\
re-sepconv & -      & -      & -          & 0.000*     & 0.095    & 0.000*  & 0.000*    & 0.833  & 0.000*  \\
soft-splat & -      & -      & -          & -          & 0.000*   & 0.354   & 0.627     & 0.000* & 0.000*  \\
ST-MFNet   & -      & -      & -          & -          & -        & 0.000*  & 0.000*    & 0.603  & 0.000*  \\
UPR-Net    & -      & -      & -          & -          & -        & -       & 0.916     & 0.000* & 0.000*  \\
VFIformer  & -      & -      & -          & -          & -        & -       & -         & 0.000* & 0.000*  \\
ACKMRF     & -      & -      & -          & -          & -        & -       & -         & -      & 0.000*  \\ \bottomrule
\end{tabular}}

\end{table}

\end{document}